\documentclass[epj]{webofc}

\usepackage{amsfonts,amsmath,amssymb,bm,hyperref}

\usepackage[varg]{txfonts}

\woctitle{QUARKS-2016: 19th International Seminar on High Energy Physics}

\begin{document}

\title{Electric current induced by an external magnetic field in the presence of electroweak matter}

\author{\firstname{Maxim} \lastname{Dvornikov}\inst{1,2}\fnsep\thanks{\email{maxdvo@izmiran.ru}}}

\institute{Pushkov Institute of Terrestrial Magnetism, Ionosphere and Radiowave Propagation (IZMIRAN), \\ Kaluzhskoe HWY 4, 108840 Troitsk, Moscow, Russia 
\and
Physics Faculty, National Research Tomsk State University, 36 Lenin
Avenue, 634050 Tomsk, Russia}

\abstract{We study the generation of an electric current, along the external magnetic field, of fermions, interacting by parity violating electroweak forces with background matter. First, we discuss the situation of massive particles with nonzero anomalous magnetic moments. We show that the induced current is vanishing for such particles in the state of equilibrium. Then, the case of massless fermions is studied. We demonstrate that the contribution of the electroweak interaction is washed out from the expression for the current, which turns out to coincide with the prediction of the chiral magnetic effect. Our results are compared with findings of other authors.}

\maketitle

\section{Introduction}

The evolution of chiral charged particles in external fields reveal
multiple quantum phenomena. First, we mention the Adler-Bell-Jackiw
anomaly, which consists in the non-conservation
of the axial current in the presence of an external electromagnetic
field. This anomaly was shown in Refs.~\cite{NieNin81,FukKhaWar08} to be closely
related to the chiral magnetic effect (the CME), which is the excitation
of the electric current of chiral particles $\mathbf{J}_{\mathrm{CME}}=\alpha_{\mathrm{em}}(\mu_{\mathrm{R}}-\mu_{\mathrm{L}})\mathbf{B}/\pi$
along the external magnetic field $\mathbf{B}$. Here $\alpha_{\mathrm{em}}\approx1/137$
is the fine structure constant and $\mu_{\mathrm{R,L}}$ are the chemical
potentials of right and left chiral fermions. One can also mention
the chiral vortical effect, which is the generation
of the anomalous current in a rotating matter. There are active searches
for manifestations of the CME in astrophysics and cosmology,
as well as in accelerator physics.

The main feature of the CME is the unbroken chiral symmetry of charged
particles. It means that any nonzero mass makes $\mathbf{J}_{\mathrm{CME}}$
to vanish~\cite{Vil80,Dvo16a}. The majority of known elementary particles
acquire masses through the electroweak mechanism. However, it is likely
to be an electroweak crossover rather than a first order phase transition.
Therefore, charged particles will remain massive at any pressure and
chemical potential unless a new physics beyond the standard model
is accounted for. There are indications that a chiral phase transition
can happen in dense matter owing to the QCD effects.
Some astrophysical applications for the magnetic fields generation
in compact stars due to the CME and the electroweak interaction between
quarks in dense matter are considered in Ref.~\cite{Dvo17b}.

In this connection, there is a particular interest in searching for
the possibility of the generation of the current $\mathbf{J}\parallel\mathbf{B}$
in the system of massive charged particles. In this situation, one
would have an instability of the magnetic field without necessity
to demand the restoration of the chiral symmetry. One of the examples
of such a system was studied in Ref.~\cite{SemSok04}, where the
dynamo amplification of magnetic fields in an inhomogeneous electroweak matter
was discussed.

There is an additional open question on the influence of the external axial-vector
field $V_{5}^{\mu}$ on the magnitude of the anomalous current in
the CME. If a homogeneous and isotropic $V_{5}^{\mu}$ is present,
the Lagrangian of the interaction of the fermion field $\psi$ with
$V_{5}^{\mu}$ can be represented as $\mathcal{L}_{5}\sim\bar{\psi}\gamma^{\mu}\gamma^{5}\psi V_{5\mu}\to\psi^{\dagger}\gamma^{5}\psi V_{5}$,
which shows that the chiral imbalance $\mu_{5}=(\mu_{\mathrm{R}}-\mu_{\mathrm{L}})/2$
could be shifted by $V_{5}\equiv V_5^0$. Thus the CME could contain the contribution
of $V_{5}$~\cite{DvoSem15a,DvoSem18,BubGubZhu17}.

In this work, we summarize our recent results in Refs.~\cite{Dvo18a,Dvo18b} on the studies of the generation of the current $\mathbf{J}\parallel\mathbf{B}$ in the presence of an axial-vector field, which can be the electroweak interaction with background matter. In Sec.~\ref{sec:AMM}, we discuss the situation of massive particles with anomalous magnetic moments. The case of massless fermions is considered in Sec.~\ref{sec:EW}. We discuss our results in Sec.~\ref{sec:CONCL}.

\section{Induced electric current of massive fermions with anomalous magnetic moments\label{sec:AMM}}

In this section, we study the generation of an equilibrium electric current of massive fermions, e.g., electrons, with anomalous magnetic moments induced by the electroweak interaction with background matter under the influence of the external magnetic field~\cite{Dvo18a}. The Lagrangian for an electron, described by the bispinor
$\psi_{e}$, has the form,
\begin{equation}\label{eq:Larg}
  \mathcal{L} = \bar{\psi}_{e}
  \left[
    \gamma_{\mu}(\mathrm{i}\partial^{\mu}+eA^{\mu}) - m +
    \frac{\mu}{2}\sigma^{\mu\nu}F_{\mu\nu} -
    \gamma_{\mu}(V_{\mathrm{L}}^{\mu}P_{\mathrm{L}} +
    V_{\mathrm{R}}^{\mu}P_{\mathrm{R}})
  \right]
  \psi_{e},
\end{equation}
where $A^{\mu}=(0,0,Bx,0)$ is the vector potential corresponding
to the constant and homogeneous magnetic field, directed along the
$z$-axis, $e>0$ is the elementary charge, $P_{\mathrm{L,R}}=(1\mp\gamma^{5})/2$
are the chiral projectors, $\gamma^{\mu}=(\gamma^{0},\bm{\gamma})$,
$\sigma_{\mu\nu}=\tfrac{\mathrm{i}}{2}[\gamma_{\mu},\gamma_{\nu}]_{-}$,
and $\gamma^{5}=\mathrm{i}\gamma^{0}\gamma^{1}\gamma^{2}\gamma^{3}$
are the Dirac matrices, $m$ is the electron mass, $\mu$ is the anomalous
magnetic moment, $F_{\mu\nu}=\partial_{\mu}A_{\nu}-\partial_{\nu}A_{\mu}=(\mathbf{E},\mathbf{B})$
is the electromagnetic field tensor (with $\mathbf{E}=0$), and $V_{\mathrm{L,R}}^{\mu}=(V_{\mathrm{L,R}}^{0},\mathbf{V}_{\mathrm{L,R}})$
are the effective potentials of the electroweak interaction of the
electron chiral projections with background matter. We shall suppose
that the background matter is macroscopically at rest and unpolarized.
In this situation, $\mathbf{V}_{\mathrm{L,R}}=0$ and $V_{\mathrm{L,R}}^{0}\equiv V_{\mathrm{L,R}}\neq0$.
The explicit form of $V_{\mathrm{L,R}}$ is given in Ref.~\cite{DvoSem15a}
for the case of background matter consisting of neutrons and protons.

Eq.~\eqref{eq:Larg} was solved in Refs.~\cite{BalStuTok13,Dvo18a}. Using the wave functions of electrons and positrons,
one gets the following expression for the total current $\mathbf{J}=\mathbf{J}_{e}+\mathbf{J}_{\bar{e}}$ along the magnetic field $\mathbf{B}$:
\begin{equation}\label{eq:Jtot}
  \mathbf{J}=\Pi\mathbf{B},
  \quad
  \Pi=-\frac{\alpha_{\mathrm{em}}}{\pi}
  \sum_{n=1}^{\infty}
  \sum_{s=\pm1}
  \int_{-\infty}^{+\infty}
  \frac{\mathrm{d}p_{z}}{\mathcal{E}}
  \left[
    p_{z}
    \left(
      1+s\frac{V_{5}^{2}}{R^{2}}
    \right) -
    s\frac{\mu BmV_{5}}{R^{2}}
  \right]
  \Delta f,
\end{equation}
where $\alpha_{\mathrm{em}}=e^{2}/4\pi$ is the fine structure constant,
$V_{5}=(V_{\mathrm{L}}-V_{\mathrm{R}})/2$, $\Delta f=f(\mathcal{E}-\chi_{\mathrm{eff}})-f(\mathcal{E}+\chi_{\mathrm{eff}})$,
$f(E)=\left[\exp(\beta E)+1\right]^{-1}$ is the Fermi-Dirac distribution
function, $\chi_{\mathrm{eff}}=\chi-\bar{V}$, $\chi$ is the chemical
potential of the electron-positron plasma, $\beta=1/T$ is the reciprocal
temperature, $\bar{V}=(V_{\mathrm{L}}+V_{\mathrm{R}})/2$, $n=1,2,\dotsc$
and $s=\pm1$ are discrete quantum numbers, which the energy levels
depend on, and
\begin{align}\label{eq:ER2}
  \mathcal{E} & =\sqrt{p_{z}^{2}+2eBn+m^{2}+(\mu B)^{2}+V_{5}^{2}+2sR^{2}},  
  \nonumber
  \\
  R^{2} & =\sqrt{(p_{z}V_{5}-\mu Bm)^{2}+2eBn
  \left[
    (\mu B)^{2}+V_{5}^{2}
  \right]}.
\end{align}
It is interesting to mention that the total current in Eq.~\eqref{eq:Jtot} is proportional to the averaged group velocity
of a charged particle along the magnetic field, $v_{z}=\partial\mathcal{E}/\partial p_{z}$:
$\mathbf{J}=-\alpha_{\mathrm{em}}\left\langle v_{z}\right\rangle \mathbf{B}/\pi$.

Considering Eq.~(\ref{eq:Jtot}) in case of a degenerate electron
gas, it was claimed in Ref.~\cite{Dvo17a} that $\Pi\neq0$. Analogous
result was obtained in Ref.~\cite{BubGubZhu17} on the basis of the
analysis of the effective Lagrangians in one-loop approximation. The instability of
the magnetic field, driven by the anomalous current $\mathbf{J}=\Pi\mathbf{B}$,
and some astrophysical applications were studied in Ref.~\cite{Dvo17a}.
The claim of Refs.~\cite{Dvo17a,BubGubZhu17} that there is $\mathbf{J}=\Pi\mathbf{B}\neq0$
in the considered system is based on the fact that the energy levels
at $n>0$ in Eq.~(\ref{eq:ER2}) are neither symmetric nor antisymmetric
functions of $p_{z}$, which is the momentum projection along the magnetic field.
Thus, the integration over $p_{z}$ in the symmetric limits in Eq.~(\ref{eq:Jtot})
could give a nonzero result. The asymmetry coefficient of the energy
levels in Eq.~(\ref{eq:ER2}) is $\mu BmV_{5}$. It is this term,
which $\mathbf{J}\parallel\mathbf{B}$ in Refs.~\cite{Dvo17a,BubGubZhu17}
is proportional to.

Nevertheless, a careful analysis reveals that $\Pi=0$ in Eq.~(\ref{eq:Jtot}).
This fact is not quite obvious. To demonstrate it, we introduce the
notation in Eq.~(\ref{eq:Jtot}),
\begin{equation}\label{eq:Fdef}
  \frac{\Delta f}{\mathcal{E}}=F(Q^{2}+2sR^{2}),
  \quad
  Q^{2}=p_{z}^{2}+2eBn+m^{2}+(\mu B)^{2}+V_{5}^{2}.
\end{equation}
Then we decompose $F(Q^{2}+2sR^{2})$,
\begin{equation}\label{eq:Fk}
  F(Q^{2}+2sR^{2}) =
  \sum_{k=0}^{\infty}
  2^{2k}R^{4k}
  \left[
    \frac{F^{(2k)}(Q^{2})}{(2k)!}+2sR^{2}\frac{F^{(2k+1)}(Q^{2})}{(2k+1)!}
  \right],
  \quad
  F^{(k)}(Q^{2}) =\frac{\mathrm{d}^{k}F(Q^{2})}{\mathrm{d}(Q^{2})^{k}},
\end{equation}
in a formal series.

The sum over $s$ of the integrand in Eq.~(\ref{eq:Jtot}) gives
\begin{align}\label{eq:I1}
  I = & \sum_{s=\pm1}
  \left[
    p_{z}+s(p_{z}V_{5}-\mu Bm)\frac{V_{5}}{R^{2}}
  \right]
  F(Q^{2}+2sR^{2})
  \nonumber
  \\
  & = 2\sum_{k=0}^{\infty}2^{2k}R^{4k}
  \left[
    p_{z}\frac{F^{(2k)}(Q^{2})}{(2k)!}+2V_{5}(p_{z}V_{5}-\mu Bm)
    \frac{F^{(2k+1)}(Q^{2})}{(2k+1)!}
  \right]
  \nonumber
  \\
  & =2\sum_{k=0}^{\infty}2^{2k}
  \left\{
    p_{z}R^{4k}\frac{F^{(2k)}(Q^{2})}{(2k)!}+
    \frac{1}{k+1}\frac{\partial}{\partial p_{z}}
    \left[
      R^{4(k+1)}
    \right]
    \frac{F^{(2k+1)}(Q^{2})}{(2k+1)!}
  \right\},
\end{align}
where we use the identities,
\begin{equation}
  \frac{\partial R^{4}}{\partial p_{z}}=2V_{5}(p_{z}V_{5}-\mu Bm),
  \quad
  \frac{1}{k+1}\frac{\partial}{\partial p_{z}}
  \left[
    R^{4(k+1)}
  \right] =
  R^{4k}\frac{\partial R^{4}}{\partial p_{z}},
\end{equation}
and take into account Eq.~(\ref{eq:ER2}).

Integrating Eq.~(\ref{eq:I1}) over $p_{z}$ and then by parts, one
gets
\begin{align}\label{eq:I2}
  \int_{-\infty}^{+\infty}I\mathrm{d}p_{z} = &
  2\sum_{k=0}^{\infty}2^{2k}
  \Bigg[
    \int_{-\infty}^{+\infty}\mathrm{d}p_{z}
    \left\{
      R^{4k}p_{z}\frac{F^{(2k)}(Q^{2})}{(2k)!} -
      \frac{1}{k+1}\frac{\partial}{\partial p_{z}}
      \left[
        F^{(2k+1)}(Q^{2})
      \right]
      \frac{R^{4(k+1)}}{(2k+1)!}
    \right\}
    \nonumber
    \\
    & +
    \left.
      \frac{F^{(2k+1)}(Q^{2})R^{4(k+1)}}{(k+1)(2k+1)!}
    \right|_{-\infty}^{+\infty}
  \Bigg].
\end{align}
The function $F(Q^{2})$ is proportional to the Fermi-Dirac distribution
functions, which are vanishing at great values of the argument. The same
property has any derivative of $F(Q^{2})$ in Eq.~(\ref{eq:Fk}).
Thus, the last term in Eq.~(\ref{eq:I2}) disappears at $p_{z}\to\pm\infty$.

Taking into account that
\begin{equation}
  \frac{\partial}{\partial p_{z}}
  \left[
    F^{(2k+1)}(Q^{2})
  \right] =
  2p_{z}\frac{\mathrm{d}}{\mathrm{d}Q^{2}}
  \left[
    F^{(2k+1)}(Q^{2})
  \right] =
  2p_{z}F^{(2k+2)}(Q^{2}),
\end{equation}
and changing the summation index $k\to k-1$ in the second term in
Eq.~(\ref{eq:I2}), one obtains
\begin{align}\label{eq:I3}
  \int_{-\infty}^{+\infty}I\mathrm{d}p_{z}= & 
  2\sum_{k=0}^{\infty}2^{2k}
  \int_{-\infty}^{+\infty}\mathrm{d}p_{z}
  R^{4k}p_{z}\frac{F^{(2k)}(Q^{2})}{(2k)!} -
  2\sum_{k=1}^{\infty}2^{2k}
  \int_{-\infty}^{+\infty}\mathrm{d}p_{z}
  R^{4k}p_{z}\frac{F^{(2k)}(Q^{2})}{(2k)!}
  \nonumber
  \\
  & =
  2\int_{-\infty}^{+\infty}\mathrm{d}p_{z} \, p_{z}F(Q^{2})=0,
\end{align}
where we use the fact that $F^{(0)}(Q^{2})=F(Q^{2})$. To get the
vanishing result of the integration in the symmetric limits in Eq.~(\ref{eq:I3}),
we account for that $Q^{2}$ is the even function of $p_{z}$ (see
Eq.~(\ref{eq:Fdef})), i.e. the integrand in Eq.~(\ref{eq:I3})
is the odd function.

Thus, we have demonstrated that in Eq.~(\ref{eq:Jtot})
\begin{equation}\label{eq:canccur}
  \mathbf{J} = \Pi \mathbf{B}=0,
  \quad
  \text{since}
  \quad
  \Pi=0,
\end{equation}
for arbitrary characteristics of the external fields and charged particles.
Accounting for the fact that the lowest energy level with $n=0$ does
not contribute to the anomalous current either~\cite{Dvo18a,Dvo17a}, we get that there is no electric current
$\mathbf{J}\parallel\mathbf{B}$ in the system of massive electrons with
anomalous magnetic moments, electroweakly interacting with background
matter, in the state of equilibrium.

\section{Induced electric current of massless fermions electroweakly interacting with background matter\label{sec:EW}}

In this section, we analyze the influence of the electroweak interaction of background matter on the generation of an electric current of massless fermions along the external magnetic field.~\cite{Dvo18b}. The Lagrangian for such a fermion, e.g., an electron, can be obtained on the basis of Eq.~\eqref{eq:Larg} if we set $\mu=0$ there. Before the consideration of a massless electron, we obtain the solution of the corresponding Dirac equation in the situation  $m\neq 0$ in Eq.~\eqref{eq:Larg}. Then, we approach to the chiral limit $m\to 0$.

Let us look for the solution of the Dirac equation, which results
from Eq.~\eqref{eq:Larg} with $\mu=0$, in the form,
\begin{equation}\label{eq:psiel}
  \psi_{e}=\exp
  \left(
    -\mathrm{i}Et+\mathrm{i}p_{y}y+\mathrm{i}p_{z}z
  \right)
  \psi_{x},
\end{equation}
where $\psi_{x}=\psi(x)$ is the bispinor which depends on $x$ and
$p_{y,z}$ are the momentum projections along the $y$- and $z$-axes.
We shall choose the chiral representation of the Dirac matrices~\cite{ItzZub80}. We can represent
$\psi_{x}$ in the form~\cite{Dvo16a},
\begin{equation}\label{eq:psix}
  \psi_{x}^{\mathrm{T}}=
  \left(
    C_{1}u_{n-1},\mathrm{i}C_{2}u_{n},C_{3}u_{n-1},\mathrm{i}C_{4}u_{n}
  \right),
\end{equation}
where $C_{i}$, $i=1,\dots,4,$ are the spin coefficients,
\begin{equation}\label{eq:Hermfun}
  u_{n}(\eta)=
  \left(
    \frac{eB}{\pi}
  \right)^{1/4}
  \exp
  \left(
    -\frac{\eta^{2}}{2}
  \right)\frac{H_{n}(\eta)}{\sqrt{2^{n}n!}},
  \quad
  n=0,1,\dotsc,
\end{equation}
are the Hermite functions, $H_{n}(\eta)$ are the Hermite polynomials,
and $\eta=\sqrt{eB}x+p_{y}/\sqrt{eB}$.

The energy spectrum for $n>0$ reads~\cite{BalPopStu11,Dvo16a}
\begin{equation}\label{eq:En>0}
  E=\bar{V}+\lambda\mathcal{E},
  \quad
  \mathcal{E}=\sqrt{(\mathcal{E}_{0}+sV_{5})^{2}+m^{2}},
  \quad
  \mathcal{E}_{0}=\sqrt{p_{z}^{2}+2eBn},
\end{equation}
where $s=\pm1$ is the discrete quantum number, dealing with the spin
operator~\cite{BalPopStu11}, and $\lambda=\pm1$
is the sign of the energy, i.e. the electron energy reads $E_{e}=E(\lambda=+1)=\mathcal{E}+\bar{V}$,
and the positron energy has the form, $E_{\bar{e}}=-E(\lambda=-1)=\mathcal{E}-\bar{V}$.
For $n=0$, one has~\cite{Dvo16a,Dvo18b}
\begin{equation}\label{eq:Elowest}
  E=\bar{V}+\lambda\mathcal{E},
  \quad
  \mathcal{E}=\sqrt{(p_{z}+V_{5})^{2}+m^{2}}.
\end{equation}
Note that, at $n=0$, there is only one spin state of the electron.

The spin coefficients obey the system~\cite{Dvo16a,Dvo18b},
\begin{align}\label{eq:Cisys}
  \left(
    \mathcal{E}\mp p_{z}\pm V_{5}
  \right)
  C_{1,3}\mp\sqrt{2eBn}C_{2,4}+mC_{3,1} & =0,  
  \nonumber
  \\
  \left(
    \mathcal{E}\pm p_{z}\pm V_{5}
  \right)
  C_{2,4}\mp\sqrt{2eBn}C_{1,3}+mC_{4,2} & =0,
\end{align}
where we consider the particle (electron) degrees of freedom, $\lambda=1$.
Since we are mainly interested in the dynamics of electrons at the
lowest energy level, we should set $n=0$ in Eq.~(\ref{eq:Cisys}).
It results from Eq.~(\ref{eq:psix}) that, in this situation, $C_{1}=C_{3}=0$
to avoid the appearance of Hermite functions with negative indexes.

If, besides setting $n=0$ in Eq.~(\ref{eq:Cisys}), we approach
to the limit $m\to0$ there, one gets
\begin{align}
  \left(
    \mathcal{E}+p_{z}+V_{5}
  \right)
  C_{2}= & 0,
  \quad
  \text{or}
  \quad
  \begin{cases}
    \mathcal{E}=-p_{z}-V_{5},
    \\
    C_{2}\neq0,
    \quad
    \text{and}
    \quad
    C_{4}=0,
  \end{cases}
  \label{eq:Rraw}
  \\
  \left(
    \mathcal{E}-p_{z}-V_{5}
  \right)
  C_{4}= & 0,
  \quad
  \text{or}
  \quad
  \begin{cases}
    \mathcal{E}=p_{z}+V_{5},
    \\
    C_{4}\neq0,
    \quad
    \text{and}
    \quad
    C_{2}=0.
  \end{cases}
  \label{eq:Lraw}
\end{align}
We can see that Eq.~(\ref{eq:Rraw}) corresponds to a right electron
and Eq.~(\ref{eq:Lraw}) to a left one.

The energy spectrum in Eq.~(\ref{eq:Elowest}) in the limit $m\to0$
reads
\begin{equation}\label{eq:chirspecgen}
  \mathcal{E}=|p_{z}+V_{5}|.
\end{equation}
Comparing Eq.~(\ref{eq:chirspecgen}) with Eqs.~(\ref{eq:Rraw})
and~(\ref{eq:Lraw}), we obtain that for a right electron
\begin{equation}\label{eq:Rrange}
  |p_{z}+V_{5}|=-p_{z}-V_{5},\quad\text{or}\quad p_{z}<-V_{5},
\end{equation}
and
\begin{equation}\label{eq:Lrange}
  |p_{z}+V_{5}|=p_{z}+V_{5},\quad\text{or}\quad p_{z}>-V_{5},
\end{equation}
for a left particle.

Therefore the total energy of a left electron at the lowest energy
level has the form,
\begin{equation}\label{eq:ELcorrect}
  E_{e\mathrm{L}}^{(n=0)}=V_{\mathrm{L}}+p_{z},
  \quad
  -V_{5}<p_{z}<+\infty,
\end{equation}
and
\begin{equation}\label{eq:ERcorrect}
  E_{e\mathrm{R}}^{(n=0)}=V_{\mathrm{R}}-p_{z},
  \quad
  -\infty<p_{z}<-V_{5},
\end{equation}
of a right particle. Comparing Eqs.~(\ref{eq:ELcorrect}) and~(\ref{eq:ERcorrect})
with analogous expressions obtained in Refs.~\cite{DvoSem15a,DvoSem18}, one can see that that the form of the spectrum at $n=0$ formally coincides with that in Refs.~\cite{DvoSem15a,DvoSem18}. However, the
range of the $p_{z}$ variation is different.

To complete the solution of the Dirac equation at $n=0$ and $m\to0$
we should fix the remaining spin coefficients. One gets that
\begin{equation}\label{eq:C24}
  C_{2}^{(\mathrm{R})}=C_{4}^{(\mathrm{L})}=\frac{1}{2\pi},
  \quad
  C_{2}^{(\mathrm{L})}=C_{4}^{(\mathrm{R})}=0,
\end{equation}
which results from the normalization condition
\begin{equation}\label{eq:psinorm}
  \int\mathrm{d}^{3}x\psi_{p_{y}p_{z}n}^{\dagger}\psi_{p'_{y}p'_{z}n'} =
  \delta
  \left(
    p_{y}-p'_{y}
  \right)
  \delta
  \left(
    p_{z}-p'_{z}
  \right)
  \delta_{nn'},
\end{equation}
of the total wave function.

The wave function of a positron can be obtained from Eqs.~(\ref{eq:psiel})
and~(\ref{eq:psix}) by applying the charge conjugation $\psi_{\bar{e}}=\mathrm{i}\gamma^{2}\psi_{e}^{*}$
and setting $\lambda=-1$ in Eq.~(\ref{eq:En>0}). Finally one has
\begin{equation}\label{eq:psipos}
  \psi_{\bar{e}}^{\mathrm{T}}= 
  \exp(-\mathrm{i}E_{\bar{e}}t-\mathrm{i}p_{y}y-\mathrm{i}p_{z}z)
  \times
  \left(
    -\mathrm{i}C_{4}u_{n},-C_{3}u_{n-1},\mathrm{i}C_{2}u_{n},C_{1}u_{n-1}
  \right),
\end{equation}
where the coefficients $C_{i}$ obey the system in Eq.~(\ref{eq:Cisys}).

If $n=0$, we obtain on the basis of Eqs.~(\ref{eq:psipos}) and~(\ref{eq:Elowest})
that
\begin{equation}\label{eq:psiposR}
  \psi_{\bar{e}\mathrm{R}}^{(n=0)}=
  \exp(-\mathrm{i}E_{\bar{e}\mathrm{R}}t-\mathrm{i}p_{y}y-\mathrm{i}p_{z}z)
  \times
  \frac{\mathrm{i}u_{0}}{2\pi}
  \left(
    -1,0,0,0
  \right)^{\mathrm{T}},
\end{equation}
where
\begin{equation}\label{eq:EposR}
  E_{\bar{e}\mathrm{R}}^{(n=0)}=p_{z}-V_{\mathrm{R}},
  \quad
  -V_{5}<p_{z}<+\infty,
\end{equation}
is the energy of right positrons at the lowest energy level. For left
positrons one has
\begin{equation}\label{eq:psiposL}
  \psi_{\bar{e}\mathrm{L}}^{(n=0)}=
  \exp(-\mathrm{i}E_{\bar{e}\mathrm{L}}t-\mathrm{i}p_{y}y-\mathrm{i}p_{z}z)
  \times
  \frac{\mathrm{i}u_{0}}{2\pi}
  \left(
    0,0,1,0
  \right)^{\mathrm{T}},
\end{equation}
where
\begin{equation}\label{eq:EposL}
  E_{\bar{e}\mathrm{L}}^{(n=0)}=-p_{z}-V_{\mathrm{L}},
  \quad
  -\infty<p_{z}<-V_{5},
\end{equation}
is the energy of left positrons at the lowest energy level. The positron
wave functions in Eqs.~(\ref{eq:psiposR}) and~(\ref{eq:psiposL})
satisfy the normalization condition in Eq.~(\ref{eq:psinorm}).

The contributions of left
and right electrons at the lowest energy level to the current are
\begin{equation}\label{eq:JLRgen}
  \mathbf{J}_{e\mathrm{L,R}}^{(n=0)}=
  -e\int\mathrm{d}p_{y}\mathrm{d}p_{z}
  \bar{\psi}_{e  \mathrm{L,R}}\bm{\gamma}\psi_{e\mathrm{L,R}}
  f(E_{e\mathrm{L,R}}^{(n=0)}-\mu_{\mathrm{L,R}}),
\end{equation}
where
$\mu_{\mathrm{L,R}}$ are the chemical potentials of left and right
particles. First we notice that the components of the current, transverse
with respect to $\mathbf{B}$, are vanishing. Performing the integration
over $-\infty<p_{y}<+\infty$ and accounting for Eqs.~(\ref{eq:ELcorrect})-(\ref{eq:C24}),
on the basis of Eq.~(\ref{eq:JLRgen}) we obtain the expression for
the total current of electrons $\mathbf{J}_{e}^{(n=0)}=\mathbf{J}_{\mathrm{L}}^{(n=0)}+\mathbf{J}_{\mathrm{R}}^{(n=0)}$
at $n=0$, 
\begin{align}\label{eq:Je}
  \mathbf{J}_{e}^{(n=0)}= &
  \frac{e^{2}\mathbf{B}}{(2\pi)^{2}}
  \left[
    \int_{-\infty}^{-V_{5}}\mathrm{d}p_{z}f(-p_{z}+V_{\mathrm{R}}-\mu_{\mathrm{R}})- 
    \int_{-V_{5}}^{+\infty}\mathrm{d}p_{z}f(p_{z}+V_{\mathrm{L}}-\mu_{\mathrm{L}})
  \right]
  \nonumber
  \\
  & =
  \frac{e^{2}\mathbf{B}}{(2\pi)^{2}}\int_{0}^{+\infty}\mathrm{d}p
  \left[
    f(p+\bar{V}-\mu_{\mathrm{R}})-f(p+\bar{V}-\mu_{\mathrm{L}})
  \right].
\end{align}
Analogously to Refs.~\cite{DvoSem15a,DvoSem18} one can show that
higher energy levels with $n>0$ do not contribute to the current.
Thus we shall omit the superscript in Eq.~(\ref{eq:Je}) for brevity.

The positron contribution to the current $\mathbf{J}_{\bar{e}}$ can
be obtained analogously to Eq.~(\ref{eq:JLRgen}) as
\begin{equation}\label{eq:JLRpos}
  \mathbf{J}_{\bar{e}\mathrm{L,R}}^{(n=0)}=
  e\int\mathrm{d}p_{y}\mathrm{d}p_{z}
  \bar{\psi}_{\bar{e}\mathrm{L,R}}\bm{\gamma}\psi_{\bar{e}\mathrm{L,R}}
  f(E_{\bar{e}\mathrm{L,R}}^{(n=0)}+\mu_{\mathrm{L,R}}).
\end{equation}
Using Eqs.~(\ref{eq:psiposR})-(\ref{eq:EposL}), we obtain on the
basis of Eq.~(\ref{eq:JLRpos}) the total contribution of positrons
at the lowest energy level to the current in the form,
\begin{align}\label{eq:Jpos}
  \mathbf{J}_{\bar{e}}^{(n=0)}= & \frac{e^{2}\mathbf{B}}{(2\pi)^{2}}
  \left[
    \int_{-\infty}^{-V_{5}}\mathrm{d}p_{z}f(-p_{z}-V_{\mathrm{L}}+\mu_{\mathrm{L}})
    -\int_{-V_{5}}^{+\infty}\mathrm{d}p_{z}f(p_{z}-V_{\mathrm{R}}+\mu_{\mathrm{R}})  
  \right]
  \nonumber
  \\
  & =
  \frac{e^{2}\mathbf{B}}{(2\pi)^{2}}\int_{0}^{+\infty}\mathrm{d}p
  \left[
    f(p-\bar{V}+\mu_{\mathrm{L}})-f(p-\bar{V}+\mu_{\mathrm{R}})
  \right].
\end{align}
It should be noted that higher energy levels do not contribute to
the current.

Using Eqs.~(\ref{eq:Je}) and~(\ref{eq:Jpos}), we obtain that the
total current $\mathbf{J}=\mathbf{J}_{e}+\mathbf{J}_{\bar{e}}$ reads
\begin{align}\label{eq:Jfinal}
  \mathbf{J}= & \frac{e^{2}\mathbf{B}}{(2\pi)^{2}}\int_{0}^{+\infty}\mathrm{d}p
  \left[
    f(p+\bar{V}-\mu_{\mathrm{R}})-f(p-\bar{V}+\mu_{\mathrm{R}})-
    f(p+\bar{V}-\mu_{\mathrm{L}})+f(p-\bar{V}+\mu_{\mathrm{L}})
  \right]
  \nonumber
  \\
  & =
  \frac{2\alpha_{\mathrm{em}}}{\pi}\mu_{5}\mathbf{B} \equiv
  \mathbf{J}_\mathrm{CME},
\end{align}
which is in agreement with the predictions of the CME at the absence of the electroweak interaction.

\section{Conclusion\label{sec:CONCL}}

In this work, we have analyzed the possibility of the existence of
the electric current induced along the external magnetic field in
the system of massive charged electrons, having anomalous magnetic
moments and electroweakly interacting with background matter, which
was supposed to be nonmoving and unpolarized. We have obtained that both the lowest, with $n=0$ (see also Refs.~\cite{Dvo17a,Dvo18a}), and higher, with $n>0$, energy levels do not contribute to this current.

The analysis of the contribution of the higher energy levels with
$n>0$ to the induced current is not trivial. Nevertheless, in Sec.~\ref{sec:AMM},
we have revealed that this contribution is vanishing; cf. Eq.~(\ref{eq:canccur}).
This result is valid at any characteristics of the electron-positron
field, such as $m$, $\mu$, etc., and any parameters of the external
fields, such $B$ and $V_{5}$.

The cancellation of the induced current $\mathbf{J}\parallel\mathbf{B}$
for $n>0$ in the considered system, which was supposed to be in the equilibrium,
corrects the recent claims in Refs.~\cite{Dvo17a,BubGubZhu17} that
such a current can be nonzero. The incorrect nonzero expression for
the current, obtained Ref.~\cite{Dvo17a}, was because of the error
in the integration over the longitudinal momentum in the case of the
degenerate electron gas. 
The discrepancy between our results and the findings of Ref.~\cite{BubGubZhu17} can be explained
by the consideration of a nonequilibrium state of the system in Ref.~\cite{BubGubZhu17}. Therefore the current $\mathbf{J} \parallel \mathbf{B} \neq 0$, derived in Ref.~\cite{BubGubZhu17}, will tend to zero very rapidly in a realistic medium.


The main reason for the cancellation of the current consists in the
fact that the longitudinal momentum $p_{z}$ can vary from $-\infty$
to $+\infty$ for a particle with a nonzero mass. Even the feature
of the energy spectrum for $n>0$ that it is not symmetric with respect
to the transformation $p_{z}\to-p_{z}$ (see Eq.~(\ref{eq:ER2})),
which Ref.~\cite{Dvo17a} appealed to in order to justify the existence
of the nonzero induced current, does not help to generate $\mathbf{J}\parallel\mathbf{B}\neq0$.
Thus the cases of $m\neq0$ and $m=0$ are different generically.
In the latter situation, the induced current can exist owing to the
CME, which is based on the asymmetric
motion of charged massless particles at the lowest energy level with
respect to the external magnetic field. The difference between the
systems of massive and massless particles consists in the chiral symmetry: it is broken in the former case and restored
in the latter one.

In the present work, we have also elaborated the improved derivation of
the anomalous current of massless charged fermions, interacting with
an axial-vector field under the influence of the external magnetic
field, flowing along the magnetic field. We have chosen a particular
example of the axial-vector field as the electroweak interaction of
an electron with nonmoving and unpolarized background matter. Unlike
Refs.~\cite{KapRedSen17,SadIsa11}, here we have used the method
of the relativistic quantum mechanics, originally proposed in Ref.~\cite{Vil80}
to describe the CME.

Using the exact solution of the Dirac equation, found in Refs.~\cite{BalPopStu11,Dvo16a,Dvo18b},
we have shown in Sec.~\ref{sec:EW} that the axial-vector field
does not contribute to the current $\mathbf{J}\parallel\mathbf{B}$;
cf. Eq.~(\ref{eq:Jfinal}). The value of the current coincides with
the prediction of the CME even in the case
when chiral fermions electroweakly interact with background matter,
confirming the findings of Refs.~\cite{KapRedSen17,SadIsa11}.

To obtain this result in frames of the relativistic quantum mechanics
one has to consider the solution of the Dirac equation for a massive
electron in the external fields and then approach to the limit $m\to0$.
If one sets $m=0$ in the Dirac equation from the very beginning,
i.e. if one considers the chiral Lagrangian in Eq.~(\ref{eq:Larg}) with $\mu = 0$ and $m=0$,
one obtains the anomalous current coinciding with that in Refs.~\cite{DvoSem15a,DvoSem18},
which is inconsistent with the results of Refs.~\cite{KapRedSen17,SadIsa11}.
Thus we conclude that the system of chiral fermions, where the external
axial-vector field is present, can be prepared in two non-equivalent
ways~\cite{Dvo18b}.

I am thankful to the organizers of Quarks-2018 for the inivitation and to RFBR (research project No.~18-02-00149a) for a partial support.


\begin{thebibliography}{50}

\bibitem{NieNin81}
  H.~B.~Nielsen and M.~Ninomiya,
  Phys. Lett. B \textbf{130}, 389 (1983).

\bibitem{FukKhaWar08}
  K.~Fukushima, D.~E.~Kharzeev, and H.~J.~Warringa,
  Phys. Rev. D \textbf{78}, 074033 (2008).

\bibitem{Vil80}
  A.~Vilenkin,
  Phys. Rev. D \textbf{22}, 3080 (1980).

\bibitem{Dvo16a}
  M.~Dvornikov,
  Phys. Lett. B \textbf{760}, 406 (2016).

\bibitem{Dvo17b}
  M.~Dvornikov,
  Int. J. Mod. Phys. D \textbf{27}, 1750184 (2018).

\bibitem{SemSok04}
  V.~B.~Semikoz and D.~D.~Sokoloff,
  Phys. Rev. Lett. \textbf{92}, 131301 (2004).

\bibitem{DvoSem15a}
  M.~Dvornikov and V.~B.~Semikoz,
  Phys. Rev. D \textbf{91}, 061301 (2015).

\bibitem{DvoSem18}
  M.~Dvornikov and V.~B.~Semikoz,
  Mod. Phys. Lett. A \textbf{33}, 1850043 (2018).

\bibitem{BubGubZhu17}
  A.~F.~Bubnov, N.~V.~Gubina, and V.~Ch.~Zhukovsky,
  Phys. Rev. D \textbf{96}, 016011 (2017).

\bibitem{Dvo18a}
  M.~Dvornikov,
  arXiv:1801.07788.

\bibitem{Dvo18b}
  M.~Dvornikov,
  Phys. Rev. D \textbf{98}, 036016 (2018).

\bibitem{BalStuTok13}
  I.~A.~Balantsev, A.~I.~Studenikin, and I.~V.~Tokarev,
  Phys. Atom. Nucl. \textbf{76}, 489 (2013).

\bibitem{Dvo17a}
  M.~Dvornikov,
  JETP Lett. \textbf{106}, 775 (2017).

\bibitem{ItzZub80}
  C.~Itzykson and J.-B.~Zuber,
  \textit{Quantum field theory}
  (McGraw-Hill, New York, 1980).

\bibitem{BalPopStu11}
  I.~A.~Balantsev, Yu.~V.~Popov, and A.~I.~Studenikin,
  J. Phys. A \textbf{44}, 255301 (2011).

\bibitem{KapRedSen17}
  D.~B.~Kaplan, S.~Reddy, and S.~Sen,
  Phys. Rev. D \textbf{96}, 016008 (2017).

\bibitem{SadIsa11}
  A.~V.~Sadofyev and M.~V.~Isachenkov,
  Phys. Lett. B \textbf{697}, 404 (2011).



\end{thebibliography}
\end{document}